\documentclass[sigconf]{acmart}
\usepackage{xcolor}
\usepackage{colortbl}

\AtBeginDocument{%
  \providecommand\BibTeX{{%
    \normalfont B\kern-0.5em{\scshape i\kern-0.25em b}\kern-0.8em\TeX}}}

\usepackage{graphicx}
\usepackage{subcaption}

\usepackage{multirow}

\renewcommand\footnotetextcopyrightpermission[1]{} 
\usepackage{cleveref}

\usepackage{caption}
\usepackage{enumitem}

\renewcommand\footnotetextcopyrightpermission[1]{} 
\begin{document}

\title{Designing Robust Quantum Neural Networks via Optimized Circuit Metrics}


\author{Walid El Maouaki\textsuperscript{1}, 
Alberto Marchisio\textsuperscript{2,3}, 
Taoufik Said\textsuperscript{1},
Muhammad Shafique\textsuperscript{2,3},
and Mohamed Bennai\textsuperscript{1}
}

\affiliation{%
  \institution{\textsuperscript{1}Quantum Physics and Magnetism Team, LPMC, Faculty of Sciences Ben M’Sik, Hassan II University of Casablanca, Morocco}
  \country{\textsuperscript{2}eBrain Lab, Division of Engineering, New York University Abu Dhabi (NYUAD), Abu Dhabi, UAE\\
  \textsuperscript{3}Center for Quantum and Topological Systems (CQTS), NYUAD Research Institute, NYUAD, Abu Dhabi, UAE}
  }
\email{walid.elmaouaki-etu@etu.univh2c.ma, alberto.marchisio@nyu.edu, taoufik.said@univh2c.ma,}
\email{muhammad.shafique@nyu.edu, mohamed.bennai@univh2c.ma}

\renewcommand{\shortauthors}{El Maouaki, et al.}

\begin{abstract}
In this study, we investigated the robustness of Quanvolutional Neural Networks (QuNNs) in comparison to their classical counterparts, Convolutional Neural Networks (CNNs), against two adversarial attacks: Fast Gradient Sign Method (FGSM) and Projected Gradient Descent (PGD), for the image classification task on both Modified National Institute of Standards and Technology (MNIST) and Fashion-MNIST (FMNIST) datasets. To enhance the robustness of QuNNs, we developed a novel methodology that utilizes three quantum circuit metrics: expressibility, entanglement capability, and controlled rotation gate selection. Our analysis shows that these metrics significantly influence data representation within the Hilbert space, thereby directly affecting QuNN robustness. We rigorously established that circuits with higher expressibility and lower entanglement capability generally exhibit enhanced robustness under adversarial conditions, particularly at low-spectrum perturbation strengths where most attacks occur. Furthermore, our findings challenge the prevailing assumption that expressibility alone dictates circuit robustness; instead, we demonstrate that the inclusion of controlled rotation gates around the Z-axis generally enhances the resilience of QuNNs. Our results demonstrate that QuNNs exhibit up to 60\% greater robustness on the MNIST dataset and 40\% on the Fashion-MNIST dataset compared to CNNs. Collectively, our work elucidates the relationship between quantum circuit metrics and robust data feature extraction, advancing the field by improving the adversarial robustness of QuNNs.
\end{abstract}




\maketitle
\pagestyle{plain}

\section{Introduction}

Machine Learning (ML) is adopted across various high-stakes domains—ranging from healthcare systems \cite{yu2018artificial} to autonomous vehicles \cite{muhammad2020deep}, and security infrastructure \cite{zhang2022artificial}. This dependency not only enhances operational capabilities but also increases reliance on software that must perform flawlessly. The danger lies in the reliance on potentially flawed AI predictions which, without robust safeguards, could lead to severe consequences for businesses and safety operations. This underscores the pressing need to prioritize robustness and ensuring these technologies are resistant to adversarial attacks \cite{hamon2020robustness}. This is particularly crucial for emerging technologies like Quantum Machine Learning (QML), and more specifically, Quanvolutional Neural Networks (QuNNs). These systems are less explored in terms of security compared to their traditional AI counterparts, highlighting a vulnerability that could jeopardize operations and safety if left unaddressed.
Research outlined in \cite{akter2023exploring} and \cite{wendlinger2024comparative} identifies vulnerabilities in both traditional and quantum machine learning models to adversarial attacks. While quantum models are often more resilient due to their distinct structures, they are not immune to adversarial attacks. This issue is compounded by the lack of comprehensive empirical evidence to back up claims about the effectiveness of the quantum circuits used. Furthermore, a study presented in \cite{liu2020vulnerability} examines how the dimensionality of Hilbert space influences vulnerability in quantum neural networks, noting that increased dimensions may reduce these networks' robustness. \textit{This made us raise an important consideration in our work: could the inherent characteristics of the Hilbert space be harnessed to enhance the security of QuNNs against such vulnerabilities?}

Our work addresses existing gaps by introducing a novel methodology that suggests that specific features of the Hilbert space could enhance the robustness of QuNNs. By exploring how characteristics of quantum circuits might influence these features, our study opens new avenues for increasing the resilience of quantum models against adversarial threats. The concept of the Hilbert space in quantum computing has been quantitatively explored by Sim et al. (2019)~\cite{sim2019expressibility}, who proposed metrics to measure how effectively a quantum circuit can address the Hilbert space. They introduced two primary measures: expressibility and entanglement capability. Subsequent research~\cite{hubregtsen2021evaluation} investigated the impact of these metrics on the classification performance of quantum neural networks, focusing primarily on clean data and self-generated datasets. Our study addresses this limitation by examining how these metrics influence the performance of QuNNs on adversarially attacked real-world datasets, specifically MNIST and Fashion-MNIST. Additionally, we expanded our investigation to analyze the role of controlled rotation gate selection in enhancing model robustness. In our work we answered emerging questions: What specific aspects or parameters of quantum circuits significantly contribute to the robustness of QuNNs? In what ways does circuit expressibility influence robustness? How does the entanglement capability of a circuit impact its robustness? Which types of single-qubit gates are most effective in preserving meaningful features for robust performance? 

This methodology examines the role of quantum circuit architecture (Ansatz) in bolstering resilience against adversarial attacks. We systematically investigate various Ansatz architectures to discern how differences in Hilbert space expressibility and entanglement capability influence the behavior of models under adversarial conditions. Our exploration not only uncovers significant insights into the impact of the Hilbert space representation on the robustness of QuNNs but also underscores the critical need for more systematic studies of these models across diverse adversarial scenarios. By analyzing how quantum-specific features can enhance security and robustness in hybrid quantum-classical systems, our research contributes to a deeper understanding of QuNNs vulnerabilities and facilitates the development of more secure quantum-based AI applications. Considering the susceptibility of image-based datasets to adversarial manipulation, we utilize the QuNNs model, which is particularly adept at processing this type of data.

Recent studies have explored various aspects of QuNNs in comparison to their classical counterparts. For instance, Sooksatra et al.~\cite{sooksatra2021evaluating} conducted a comparative analysis between QuNNs and analogous classical networks, while Zaman et al.~\cite{zaman2024studying} investigated the influence of quantum hyperparameters on QuNNs accuracy. These works, however, did not encompass a comprehensive range of perturbation strength values in their evaluations.
Our research extends beyond these previous studies by examining the robustness of QuNNs models across a broader spectrum of perturbation strengths. This approach allows for a more thorough and nuanced comparison between quantum and classical neural networks.
Additionally, while Huang et al.~\cite{huang2023image} analyzed the effects of kernel size and filter number variations in QuNNs, we argue that this approach, being inspired by classical methods, does not fully capture the unique quantum dynamics at play. Our work takes a more fundamentally quantum-oriented perspective by investigating the impact of quantum circuit design on QuNNs performance.

In this paper, we present a methodology for robust QuNNs against adversarial attacks. Our contributions to this paper are:\vspace{-0.5em}

\begin{itemize}
    \item We design a methodology to robustify QuNNs against different types of white box adversarial attacks through the quantum circuit architecture.
    \item We compare the robustness of QuNNs to classical CNNs against various types of adversarial attacks. 
    \item We identify that expressibility, entanglement capability, and the selection of controlled rotation gates significantly impact QuNN robustness. Expressibility enhances resistance to low-spectrum adversarial perturbations, while lower entanglement capabilities improve robustness, indicating that simpler quantum states are more attack-resistant. Controlled rotation gates around the Z-axis effectively preserve data features critical for robust performance.

\end{itemize}

\section{Background}
\subsection{Convolutional and Quanvolutional neural networks}
Convolutional Neural Networks (CNNs)~\cite{li2021survey}, ~\cite{alzubaidi2021review}, have become an essential component of current image identification tasks due to their unique architecture, which is capable of capturing spatial structure in data. A CNN architecture typically consists of a sequence of layers: convolutional layers, pooling layers, and fully connected layers. The convolutional layer, the fundamental building block, applies a set of learnable filters, or kernels, to the input image to generate feature maps. Mathematically, the convolution operation is defined as $(X * K)[i, j]=\sum_m \sum_n X[i+m, j+n] K[m, n]$, where $X$ is the input image, $K$ is the kernel, and $*$ denotes the convolution operation. This process extracts features such as edges, textures, and shapes. Pooling layers, such as max-pooling or average-pooling, are then used to downsample the feature maps, reducing the spatial dimensions and the computational load while retaining the most significant features. \textit{In this study, we concentrate specifically on the convolutional layer in our models and deliberately omit the pooling layer. This decision allows us to delve deeper into the convolutional layer's unique contributions to feature extraction and model robustness}. 

QuNNs represent an innovative hybrid architecture that leverages quantum computing to enhance classical CNNs by incorporating quantum circuits to process data~\cite{henderson2020quanvolutional}. In a QuNN, the convolutional layers are replaced by quantum circuits, specifically designed to exploit quantum superposition and entanglement, enabling the extraction of complex features that might be challenging for classical CNNs. The quantum convolutional layer, also known as the quanvolutional layer, functions similarly to a classical convolutional layer but utilizes quantum circuits to process image patches. For an $m\times m$ input image, an $n\times n$ kernel, and a stride of $2$, the quanvolutional layer processes data using a $n^2$ qubits quantum circuit. The dimensionality of the feature maps is directly influenced by the number of qubits used, with each qubit contributing to a distinct channel in the output. QuNN process data as follows:

\textbf{Extracting patches:} Each $n \times n$ patch $P_{i, j}$ of the image $X$ is defined by:
$$
P_{i, j}=\{X[i+k, j+l] \mid 0 \leq k<n, 0 \leq l<n\}
$$
where $i$ and $j$ denote the coordinates of the top-left corner of the patch.

\textbf{Encoding patches into quantum states:} Each $n \times n$ patch $P_{i, j}$ is encoded into a quantum state $\left|\psi_{i, j}\right\rangle$. One common encoding method is angle encoding, where it maps each pixel value from the patch to the corresponding qubit using a rotation gate:
$$
\left|\psi_{i, j}\right\rangle=U_{\mathrm{enc}}\left(P_{i, j}\right)|0\rangle^{\otimes n^2}
$$
where $U_{\text {enc }}$ is the unitary operator for encoding the patch into a quantum state over $n^2$ qubits. The rotation is preferred to be in the range of $[0,1]$ as quantum rotations beyond this interval introduce periodicity redundancies (e.g., angles differing by $2\pi$ yield identical states), and normalized pixel values in $[0,1]$ ensure efficient use of the qubit state space while aligning with typical hardware constraints.

\textbf{Applying the quanvolutional filter (quantum kernel):} The encoded quantum state $\left|\psi_{i, j}\right\rangle$ is processed by an $n^2$-qubit operation $U(\theta)$ which acts as an $n \times n$ kernel:
$$
\left|\phi_{i, j}\right\rangle=U(\theta)\left|\psi_{i, j}\right\rangle
$$
Here, $U(\theta)$ is a unitary transformation designed to perform the quantum equivalent of convolution. $U(\theta)$ is a parameterized quantum circuit (PQC) which can vary depending on the gate selection and entanglement pattern chosen.

\textbf{Measurement:} Each qubit in the quantum state $\left|\phi_{i, j}\right\rangle$ is measured, and its expectation value contributes to a specific channel in the feature map. For an $n \times n$ kernel with $n^2$ qubits, the expectation value of each qubit $k$ is computed as:
$$
M_{i, j, k}=\left\langle\phi_{i, j}\left|Z_k\right| \phi_{i, j}\right\rangle
$$
where $Z_k$ is the Pauli-Z operator for the $k$-th qubit. This measurement yields a scalar value that corresponds to one pixel in the channel $k$.

\textbf{Constructing the feature map:} The feature map is constructed by applying these steps to each $n \times n$ patch, with a stride of 2. The resulting feature map $F$ consists of multiple channels $F_k$, each corresponding to one of the qubits. Formally, the feature map can be described as:
$$
F_k[i, j]=M_{i, j, k}
$$
where $k$ ranges from 1 to $n^2$, representing the different channels, and $i$ and $j$ vary in steps of 2 , i.e., $i, j \in\{0,2,4, \ldots, m-n\}$.

\textbf{Quanvolutional operation:} The quanvolutional layer's operation can be succinctly represented as:
$$F_k[i, j] = \left\langle U(\theta) U_{\mathrm{enc}}\left(P_{i, j}\right) \left| 0 \right\rangle^{\otimes n^2} \middle| Z_k \middle| U(\theta) U_{\mathrm{enc}}\left(P_{i, j}\right) \left| 0 \right\rangle^{\otimes n^2} \right\rangle
$$

\begin{figure*}[h]
    \centering
    \includegraphics[width=\linewidth]{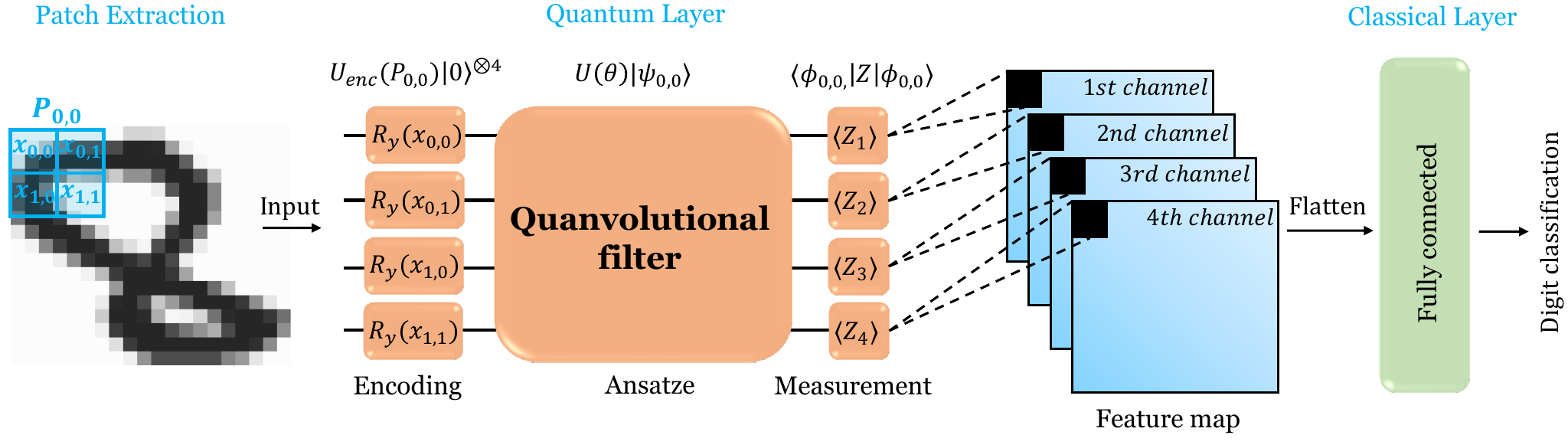}
    \caption{
    Depiction of a $2\times 2$ kernel QuNN operating on the digit $'8'$ from the MNIST collection. The process begins by extracting a $2 \times 2$ patch from the input image. Each value in this patch is encoded into a quantum state via a rotational gate around the $y$-axis. Following this, a Quanvolutional filter implements a parameterized quantum circuit (Ansatz) on the quantum states. Measurements of the output generate feature maps in various channels. These maps are subsequently flattened and directed through a fully connected classical layer for the classification of the digit.}
    \label{fig:QNN_architecture}
\end{figure*}

Finally, the feature maps produced by either the CNN or the QuNN are flattened and passed through one or more fully connected layers, which perform classification using the learned features. Refer to figure \ref{fig:QNN_architecture} for an illustration of a QuNN applied to the digit $'8'$ from the MNIST dataset.

\subsection{Quantum machine learning vulnerability}
The susceptibility of quantum classifiers to adversarial attacks has gained notable attention. Research such as~\cite{liu2020vulnerability},~\cite{el2024advqunn},~\cite{el2024robqunns} and~\cite{lu2020quantum} showcased that QML models, much like classical models, can be vulnerable to adversarial examples. They have shown that even when these quantum classifiers are implemented on noisy quantum devices, they can still be deceived by these subtle alterations to the input. As a defense strategy, one proposed technique is adversarial training, which involves including these adversarial examples during the training process to improve the model's resilience.

Building on this empirical studies, there has been progress in theoretical approaches to enhancing the adversarial robustness of QML models. For instance, this study~\cite{georgiou2024adversarial} established new information-theoretic upper bounds on the generalization error of adversarially trained quantum classifiers, establishing a crucial link between adversarial perturbations and increased sample complexity. Another significant contribution in this field employed quantum many-body physics principles to offer provable defenses for QML models~\cite{dowling2024adversarial}. This research demonstrated that quantum classifiers can inherently resist low perturbations, local attacks under non-scrambling conditions, and universal adversarial attacks in scenarios characterized by quantum chaos, fundamentally strengthening the security of QML algorithms against sophisticated adversarial attacks. These insights help illuminate the underlying dynamics of QML models under adversarial conditions, offering a pathway toward the development of more robust QML systems. 

Moreover, incorporating quantum phenomena into defense strategies is emerging as a novel method to protect QML models. Recent studies propose using quantum noise and other unique quantum properties to enhance adversarial robustness, potentially improving theoretical limits on adversarial errors~\cite{du2021quantum},~\cite{huang2023certified},~\cite{gong2024enhancing},~\cite{west2024drastic}.

\section{Preliminaries}
\subsection{Expressibility of Quantum Circuits}
In the realm of QML, the expressibility of quantum circuits is a fundamental property that determines the ability of a quantum neural network to represent complex functions. High expressibility implies that the quantum circuit can approximate any function with high fidelity, which is crucial for effective learning and generalization in QML models. Expressibility in the context of quantum circuits refers to the capability of a parameterized quantum circuit (PQC), we refer to it interchangeably as Ansatz, to produce a wide range of quantum states. Formally, let $\mathcal{U}(\theta)$ denote an Ansatz where $\theta \in \mathbb{R}^d$ is a vector of parameters. The expressibility of $\mathcal{U}(\theta)$ is quantified by how uniformly the states $\mathcal{U}(\theta)|0\rangle$ cover the unitary space with respect to the Haar measure on the unitary group $\mathcal{U}\left(2^n\right)$, where $n$ is the number of qubits.

To evaluate the expressibility of our quantum circuit, we first use fidelity as a fundamental measure of closeness between states, defined as $F(\rho, \sigma)=(\operatorname{Tr} \sqrt{\sqrt{\rho} \sigma \sqrt{\rho}})^2$. For pure states $|\psi\rangle$ and $|\phi\rangle$, this simplifies to $|\langle\psi \mid \phi\rangle|^2$. By sampling a large number of random parameters $\theta$ for the circuit $\mathcal{U}(\theta)$ and computing these fidelities, we obtain an empirical distribution $P_A(F)$. Next, we compare $P_A(F)$ to the corresponding distribution $P_H(F)$ produced by Haar-random unitaries, an "ideal" reference that represents maximum expressibility. Finally, we use the Kullback-Leibler (KL) divergence $D_{\mathrm{KL}}\left(P_A \| P_H\right)$ (as defined in Equation \ref{Kl_eq}) to quantify how closely our fidelity distribution matches that ideal. A lower KL divergence thus indicates higher expressibility, as it signals that $\mathcal{U}(\theta)$ more closely emulates Haar-random sampling~\cite{sim2019expressibility}.

\begin{equation}
    D_{K L}(P_A \| P_H)=\sum_F P_A(F) \log \frac{P_A(F)}{P_H(F)}
   \label{Kl_eq}
\end{equation}

In other words, a lower value of $D_{K L}$ indicates that the distribution $P_A(F)$ is close to $P_H(F)$, implying that our parameterized quantum circuit has high expressibility. Conversely, a higher KL divergence indicates that the circuit is less expressive.

As an example, consider a quantum circuit composed of $L$ layers, each containing a set of parameterized single qubit rotations $R_y\left(\theta_i\right)=e^{-i \theta_i Y / 2}$ and entangling gates such as CNOTs. The total unitary operation can be written as:
$$
\mathcal{U}(\theta)=\prod_{l=1}^L\left(\bigotimes_{j=1}^n R_y\left(\theta_{l j}\right) \cdot \mathcal{E}_l\right)
$$
where $\mathcal{E}_l$ denotes the entangling operations in layer $l$. We sample a large number of parameter sets $\left\{\theta_i\right\}_{i=1}^N$, and for each $\theta_i$, we compute the resulting quantum state $\left|\psi\left(\theta_i\right)\right\rangle=\mathcal{U}\left(\theta_i\right)|0\rangle$. Next, we evaluate the fidelity between pairs of states $F\left(\left|\psi\left(\theta_i\right)\right\rangle,\left|\psi\left(\theta_j\right)\right\rangle\right)$ for a subset of pairs $\left(\theta_i, \theta_j\right)$. Then we construct the empirical distribution $P_A(F)$ of these fidelities and compare it to the Haar-random distribution $P_H(F)$, which is computed by its analytical expression $(N-1)(1-F)^{N-2}$, $N=2^n$ correspond to the Hilbert space dimension, $n$ is the number of qubits. A KL divergence close to zero indicates high expressibility, while a larger divergence suggests that the circuit is less capable of uniformly exploring the quantum state space.

\begin{figure*}[htb]
  \centering
  \includegraphics[width=\textwidth]{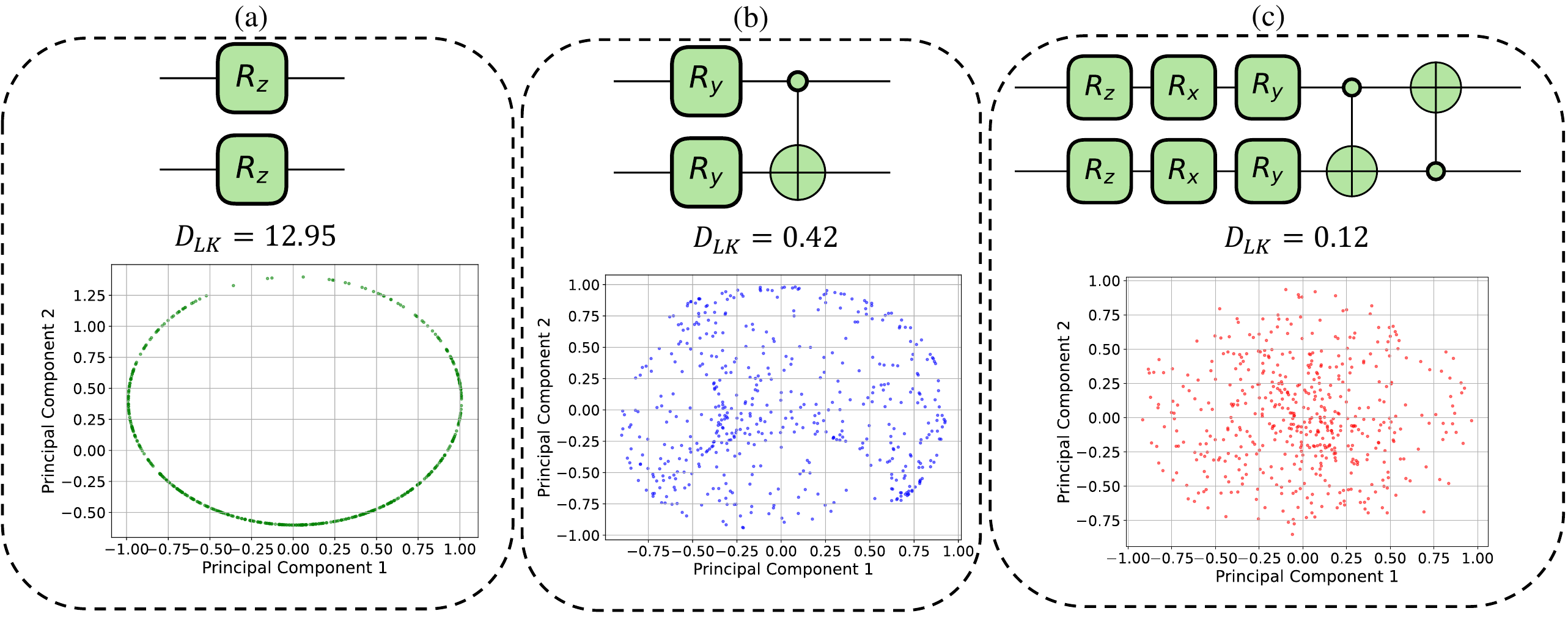}
  \caption{Visualization of quantum state space exploration: (a) Low expressibility circuit with RZ gates shows an elliptical distribution, (b) Medium expressibility circuit with RY and CNOT gates displays moderate dispersion, (c) High expressibility circuit using RX, RY, RZ, and CNOT gates demonstrates widespread, uniform state exploration.}
  \label{fig:express_example}
\end{figure*}

Figure \ref{fig:express_example} shows an example of three circuits and their exploration of spaces based on expressibility. To visualize the quantum state coverage in these spaces, we applied Principal Component Analysis (PCA) to reduce the complex, high-dimensional quantum states into a more manageable two-dimensional representation. For a two-qubit system, the state vectors exist in a 4-dimensional complex Hilbert space. By separating the real and imaginary parts, the state vectors are transformed into an 8-dimensional real space, which is then reduced to 2 dimensions using PCA. This process provides a visual representation of the circuit’s ability to explore the quantum state space.

Figure \ref{fig:express_example} a) shows a low expressibility circuit, this circuit uses only parameterized RZ gates for each qubit. RZ gates perform rotations around the Z-axis of the Bloch sphere, affecting only the phase of the qubit states. Since there are no entangling gates or rotations around other axes, 
the PCA plot shows the data points forming an elliptical distribution which is a limited form that indicates that the circuit is exploring only a restricted subset of the Hilbert space.

Figure \ref{fig:express_example} b) shows a medium expressibility circuit that applies RY gates to each qubit followed by a layer of CNOT gates that introduce entanglement between neighboring qubits. RY gates rotate the qubits around the Y-axis, allowing changes to both the phase and amplitude of the qubit states. The addition of CNOT gates introduces entanglement, increasing the circuit's ability to explore more complex quantum states. 
The PCA plot for this circuit shows a more dispersed distribution, indicating moderate exploration of the Hilbert space.

Figure \ref{fig:express_example} c) shows a high expressibility circuit, this circuit consists of a single layer that includes RX, RY, and RZ gates for each qubit, along with CNOT gates arranged in a ring topology. The combination of RX, RY, and RZ gates allows for comprehensive state rotations, altering both amplitude and phase, while the CNOT gates create entanglement among all qubits. The PCA plot for this circuit displays a widespread and uniform distribution, signifying that the circuit is capable of exploring a broad range of quantum states.

\subsection{Entanglement capability of quantum circuits}
Entanglement is a fundamental resource in quantum computing, essential for enabling the complex correlations that give quantum algorithms their advantage over classical counterparts. The entangling capability of a quantum circuit refers to its ability to generate and maintain these entangled states, which serve as another measure of the Ansatz characteristics. To quantify this capability, the Meyer-Wallach measure $Q$ is commonly used, defined as 
\begin{equation}
    Q(|\psi\rangle)=\frac{4}{n} \sum_{i=1}^n\left(1-\operatorname{Tr}\left(\rho_i^2\right)\right)
\end{equation}

where $\rho_i=\operatorname{Tr}_{\{1, \ldots, n\} \backslash i}(|\psi\rangle\langle\psi|)$ is the reduced density matrix of the $i$-th qubit obtained by tracing out all other qubits. The entangling capability of an Ansatz $\mathcal{U}(\theta)$ is then defined as the average Meyer-Wallach measure over many states generated by different parameter sets $\theta$ sampled from the distribution $\mathcal{D}$ 
\begin{equation}
    \bar{Q}=\mathbb{A}_{\theta \sim \mathcal{D}}[Q(|\psi(\theta)\rangle)]
\end{equation}
where $|\psi(\theta)\rangle=\mathcal{U}(\theta)|0\rangle^{\otimes n}$ represents the quantum state generated by the circuit for a given parameter set $\theta$. This average is taken over states randomly sampled from the circuit, providing a measure of how effectively the Ansatz can create entangled states.

For instance, we start by sampling a set of parameters $\left\{\theta_j\right\}_{j=1}^m$ from a chosen distribution $\mathcal{D}$, where each $\theta_j$ corresponds to a specific configuration of the circuit. For each sampled $\theta_j$, we compute the quantum state $\left|\psi\left(\theta_j\right)\right\rangle=$ $\mathcal{U}\left(\theta_j\right)|0\rangle^{\otimes n}$ and determine the reduced density matrices $\rho_i\left(\theta_j\right)$ for each qubit $i$. The Meyer-Wallach measure $Q\left(\left|\psi\left(\theta_j\right)\right\rangle\right)$ is then calculated for each state, and the entangling capability $\bar{Q}$ is obtained by averaging these values over all sampled parameters: $$\bar{Q}=\frac{1}{m} \sum_{j=1}^m Q\left(\left|\psi\left(\theta_j\right)\right\rangle\right)$$

A higher value of $\bar{Q}$ indicates stronger entangling capability, while a lower value suggests weaker entangling capability.

\section{Methodology}
\label{sec:methodology}

This section delineates our novel methodology which is employed to augment the robustness of QuNNs against adversarial attacks by leveraging the Hilbert space representation of the Ansatz. Our primary focus is on enhancing the resilience of QuNNs by exploring the expressibility and entanglement capabilities of quantum circuits configured through various unitary operations. The Ansatz $U$, a $2^n \times 2^n$ matrix, configures an $n$-qubit quantum circuit. This configuration involves single-qubit gates (e.g., $R_x, R_{y,} R_z, H$ ) and two-qubit gates (e.g., $C N O T, C R_x, C R_y, C R_z$ ). Such arrangements facilitate the creation of diverse Ansatzes or quanvolutional filters, each characterized by specific gate sequences and entanglement patterns. These configurations influence key metrics of the Ansatz: expressibility, entanglement capability, and gate selection, which are central to our investigation.

\begin{figure*}[htb]
  \centering
  \includegraphics[width=\textwidth]{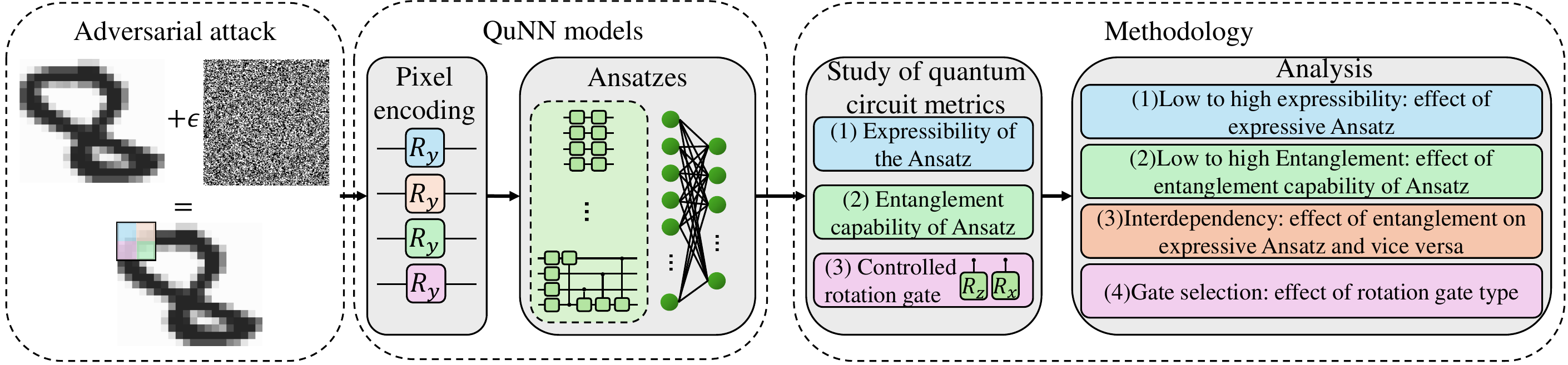}
  \caption{Overview of our proposed methodology.}
  \label{fig:methodology}
\end{figure*}

To systematically evaluate the impact of these metrics on QuNNs robustness, we conducted the following experiments, see Fig \ref{fig:methodology} for an overview of the methodology:

\begin{enumerate}
	\item \textbf{Gradational Analysis of Expressibility and Entanglement:} We aimed to examine the influence of varying levels of expressibility and entanglement capability on QuNN robustness under different intensities of adversarial perturbations. This was achieved by selecting five quantum circuits, each representing a point along a spectrum from low to high expressibility and entanglement capability.

	\item \textbf{Interdependency of Expressibility and Entanglement:} The experiment was designed to investigate how entanglement capability impacts expressive circuits and vice versa. We analyzed circuits categorized into four groups—low entanglement with low expressibility, high entanglement with low expressibility, low entanglement with high expressibility, and high entanglement with high expressibility—to understand their interdependencies and their combined effects on the robustness of QuNNs.

	\item \textbf{Gate Selection Impact:} The focus of this experiment was to assess the effects of controlled $X$ and $Z$ rotation gates on the 
    QuNN robustness. By comparing paired quantum circuits that shared the same architectural design but differed in their controlled rotation gates ($CR_x$ versus $CR_z$), we analyzed how these variations affect the network's resilience to adversarial attacks.
\end{enumerate}

Our overarching aim is to determine how different Ansatz metrics—expressibility, entanglement capability, and gate selection—impact data representation within the Hilbert space, shape the construction of decision boundaries, and affect the resilience of QuNNs against adversarial attacks. The robustness of QuNNs was evaluated using two advanced white-box adversarial attack techniques: Fast Gradient Sign Method (FGSM) and Projected Gradient Descent (PGD), which are detailed in subsequent sections.

\begin{table}[h!]
\centering
\begin{tabular}{|c|c|c|c|}
\hline
\textbf{Ansatz} & \textbf{Express $\downarrow$} & \textbf{Ent Capability $\uparrow$} & \textbf{Control gate} \\ \hline
Ansatz 1  & 0.4015 & 0.0 & $-$\\ \hline
Ansatz 2  & 0.3028 & 0.6130 & $-$\\ \hline
Ansatz 3  & 0.2536 & 0.1767 & $CRz$\\ \hline
Ansatz 4  & 0.0987 & 0.3016 & $CRx$ \\ \hline
Ansatz 5  & 0.0755 & 0.2833 & $CRz$ \\ \hline
Ansatz 6  & 0.0105 & 0.6800 & $CRx$ \\ \hline
Ansatz 7  & 0.1305 & 0.1995 & $CRz$ \\ \hline
Ansatz 8  & 0.0835 & 0.2856 & $CRx$ \\ \hline
Ansatz 9  & 0.6946 & 1.0000 & $-$ \\ \hline
Ansatz 11  & 0.1339 & 0.5336 & $CRz$ \\ \hline
Ansatz 12  & 0.2247 & 0.3999 & $CRx$ \\ \hline
Ansatz 13 & 0.0956 & 0.4078 & $CRz$ \\ \hline
Ansatz 14 & 0.0496 & 0.5403 & $CRx$ \\ \hline
Ansatz 15 & 0.1915 & 0.7094 & $-$ \\ \hline
Ansatz 16  & 0.3083 & 0.1723 & $CRz$ \\ \hline
Ansatz 17  & 0.1665 & 0.2912 & $CRx$ \\ \hline
Ansatz 18  & 0.2495 & 0.2100 & $CRz$ \\ \hline
Ansatz 19 & 0.0933 & 0.3958 & $CRx$ \\ \hline
\end{tabular}
\caption{Quantum metrics of Ansatzes: Expressibility, Entanglement Capability and controlled rotation gate of Selected Ansatz Circuits.}
\label{table:ansatz}
\end{table}

\begin{table*}[t]
\centering
\begin{tabular}{|c|c|c|c|}
\hline
\textbf{Category} & \textbf{Ansatz} & \textbf{Exprssibility $\downarrow$} & \textbf{Entanglement $\uparrow$} \\ \hline
\multirow{5}{*}{Low to High Expressibility} 
& Ansatz 9  & 0.6946 & $-$ \\ \cline{2-4} 
& Ansatz 15 & 0.1915 & $-$ \\ \cline{2-4} 
& Ansatz 13 & 0.0956 & $-$ \\ \cline{2-4} 
& Ansatz 14 & 0.0496 & $-$ \\ \cline{2-4} 
& Ansatz 6  & 0.0105 & $-$ \\ \hline
\multirow{5}{*}{Low to High Entanglement} 
& Ansatz 1  & $-$ & 0.0  \\ \cline{2-4} 
& Ansatz 3  & $-$ & 0.1767 \\ \cline{2-4} 
& Ansatz 19 & $-$ & 0.3958 \\ \cline{2-4} 
& Ansatz 2  & $-$ & 0.6130 \\ \cline{2-4} 
& Ansatz 9  & $-$ & 1.0000  \\ \hline
\multirow{4}{*}{Low-High Combination} 
& Low-Ent, Low-Exp (Ansatz 1)   & 0.4015 & 0.0 \\ \cline{2-4} 
& High-Ent, Low-Exp (Ansatz 9)  & 0.6946 & 1.0  \\ \cline{2-4} 
& Low-Ent, High-Exp (Ansatz 3)  & 0.2536 & 0.1767  \\ \cline{2-4} 
& High-Ent, High-Exp (Ansatz 6) & 0.0105 & 0.6800 \\ \hline
\end{tabular}%
\caption{Expressibility and Entanglement Capability of selected Ansatz circuits in our methodology.}
\label{table:ansatz_2}
\end{table*}

\subsection{Adversarial attacks on QuNNs and CNNs} \label{ADVattack}
Given a machine learning model, whether classical or quantum, we train it iteratively until it converges to a local optimal solution, achieving a satisfactory separation of the dataset.

In a white-box adversarial attack scenario, the adversary has full knowledge of the model, including its architecture, parameters, and gradients. This knowledge is exploited to craft adversarial examples that aim to mislead CNN into making incorrect classifications. The following methods are employed to generate these adversarial examples:

\begin{enumerate}
    \item Fast Gradient Sign Method (FGSM) \cite{goodfellow2014explaining}:
 is a gradient based attack that perturbs the input image $x$ by adding a small perturbation $\epsilon$ in the direction of the gradient of the loss function $J(\theta, x, y)$ with respect to the input, where $\theta$ represents the model parameters, and $y$ is the true label. The adversarial example $x^{\prime}$ is computed as:
$$
x^{\prime}=x+\epsilon \cdot \operatorname{sign}\left(\nabla_x J(\theta, x, y)\right)
$$

Here, $\epsilon$ is a small scalar value that controls the strengths of the perturbation.
    \item Projected Gradient Descent (PGD) \cite{madry2018towards}:
 is an iterative extension of FGSM. It applies multiple iterations of small FGSM like steps and projecting the perturbed image back into the feasible input space. The adversarial example after $k$ iterations is given by:
$$
x_{k+1}=\operatorname{Proj}_\epsilon\left(x_k+\alpha \cdot \operatorname{sign}\left(\nabla_x J\left(\theta, x_k, y\right)\right)\right)
$$
where $\alpha$ is the step size, and $\operatorname{Proj}_\epsilon$ denotes the projection operator that ensures $x_{k+1}$ remains within the $\epsilon$-ball around the original input $x$.

\end{enumerate}

\section{Results and Discussion}

\begin{figure}
    \centering
    \includegraphics[width=\linewidth]{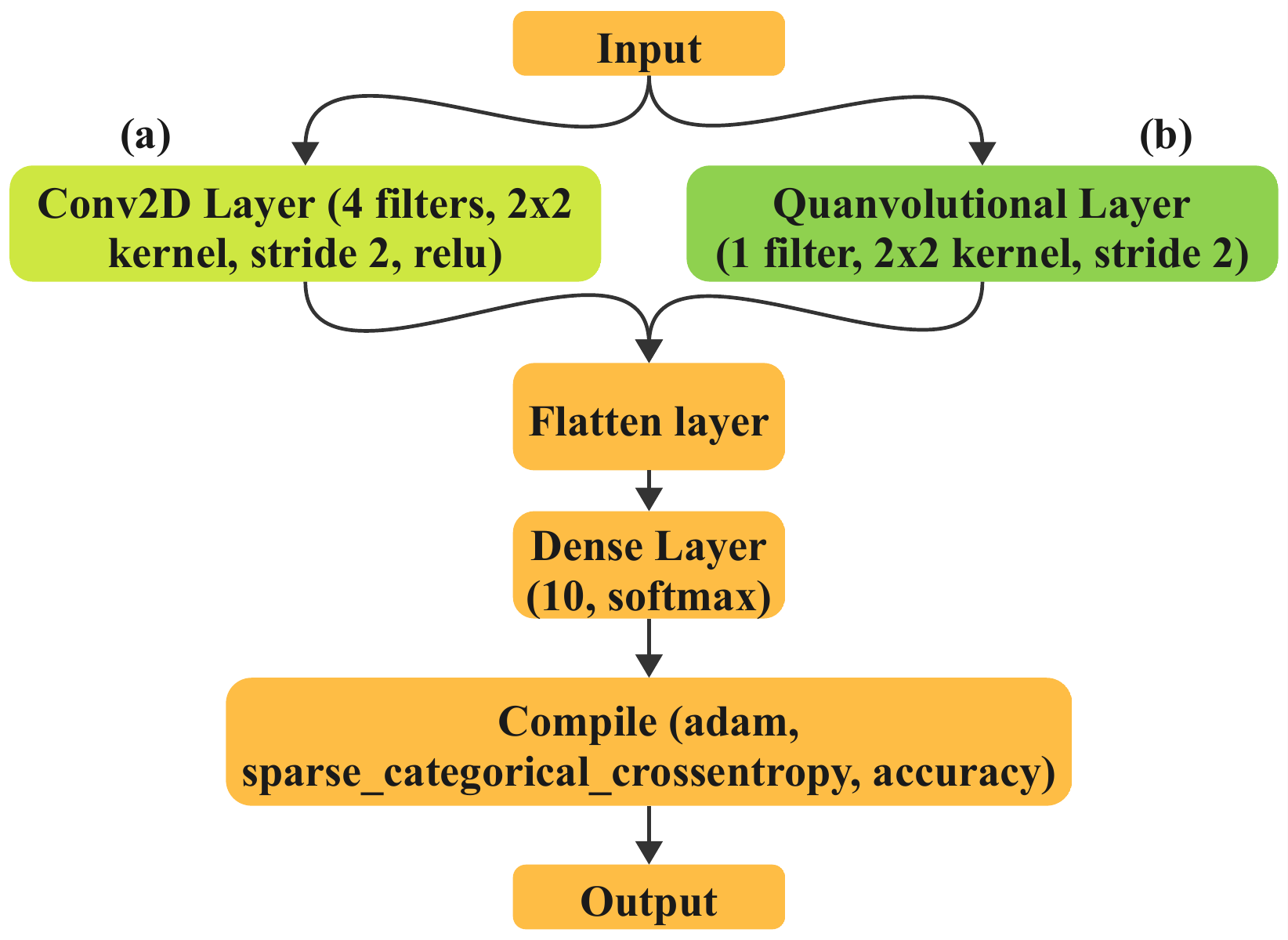}
    \caption{Diagram illustrating (a) a CNN featuring a convolutional layer, and (b) a QuNN equipped with a quantum convolutional layer, both applied to MNIST and FMNIST datasets. After flattening, the data proceeds to the softmax layer, followed by optimization of the model and evaluation of its performance.}
    \label{fig:Flows}
\end{figure}

\subsection{Experimental Setting}
The adversarial robustness of QuNNs and CNNs was assessed using two datasets MNIST dataset~\cite{mnist} and FMNIST dataset~\cite{FashionMNIST}, consisting of 28x28 pixel grayscale images, against two adversarial attack methods: FGSM and PGD, each at varying levels of perturbation strength. See Figure \ref{fig:Flows} for an overview of the quantum and classical model architectures. The data was normalized to be in the range $[0,1]$ before training and evaluation for both quantum and classical models. This step ensures compliance with the input limits of quantum circuits, where rotation angles must remain within set boundaries, and also aids in stabilizing training in traditional CNNs by avoiding excessive activations.

The CNN model utilizes a classical Conv2D layer with 4 filters and uses the ReLU activation function to introduce non-linearity. In contrast, QuNNs employ a quantum convolutional (Quanvolutional) layer, which uses a single filter corresponding to one quantum circuit layer and a 2x2 kernel representing 4 qubits in the circuit, with a stride of 2. Both models produce feature maps with dimensions of 14x14 pixels and 4 channels.
After convolution, each model flattens the resulting feature maps into a one-dimensional vector, which is then fed into a dense layer with 10 neurons. This layer uses the softmax activation function to output a probability distribution across 10 classes (digits 0-9). Each model is compiled using the Adam optimizer, with sparse categorical cross-entropy as the loss function and accuracy as the performance metric.
For this study, in both models, the convolutional layers act as fixed feature extractors for both classical and quantum models, meaning their parameters are non-trainable. 
However, the fully connected layer is trainable, enabling model optimization.

In our experimental design, we aimed to isolate the influence of the Ansatz in the Quanvolutional layer on the robustness of QuNNs. To achieve this, we required an embedding method with minimal expressive power and no entanglement capability. After evaluating various quantum embedding techniques, we selected angle encoding (a single parameterized Pauli-Y rotation per qubit) for its simplicity and minimal impact on circuit metrics. This choice is consistent with prior work demonstrating the suitability of angle encoding for QuNNs \cite{henderson2020quanvolutional}, \cite{mattern2021variational}. By using this straightforward rotational encoding, we minimized additional complexities—such as deeper circuits, non-zero L2-norm requirements, handling zero-valued patches, and unintended data rescaling—that could arise from alternative encoding schemes like amplitude encoding and potentially obscure the Ansatz’s impact. By maintaining low expressibility and zero entanglement with angle encoding, we ensure that our focus remains solely on how the Quanvolutional Ansatz influences the robustness of the QuNNs.

The models are trained using $1000$ samples from each dataset over $30$ epochs, with a batch size of four, and a learning rate of $0.001$. We use these trained models for white-box attacks to generate adversarial examples using different attack methods: FGSM and PGD. The adversarial examples are generated using the TensorFlow Cleverhans library, which computes the gradient of the loss function on the image pixels and updates the adversarial examples as described in subsection \ref{ADVattack}. We vary the perturbation strength from low to high relative to each attack type (FGSM: low 0–0.4, high 0.6–1; PGD: low 0–0.04, high 0.06–0.1), and create a set of adversarial examples for each level of perturbation. Lastly, the performance of the trained models is assessed using the adversarial images across the range of epsilon values, which produces plots of accuracy against epsilon. To enhance the reliability of our findings and to control for variability in weight initialization, we conducted 10 independent runs for each experiment. The mean values were calculated and plotted along with error bars to illustrate the statistical variability of the results.

The Ansatz selections utilized in this study are derived from work referenced in~\cite{sim2019expressibility}. A detailed view of the quantum circuits can be found in Appendix A, Figure \ref{fig:gate_selection}. Table \ref{table:ansatz} comprehensively illustrates all the Ansatzes employed in our experiments, complete with their corresponding expressibility, entanglement capability values, and the controlled rotation gate used. Figure \ref{fig:Spectrum} in Appendix A presents a comprehensive diagram encompassing all the measures outlined in Table \ref{table:ansatz}. Following this overview, we present the results of our experiments, which explore the impact of these metrics on the robustness of QuNNs.

\subsection{QuNN versus CNN}
The study revealed that the QuNN models exhibited superior resistance to adversarial attacks when compared to traditional CNN architectures. Figure \ref{fig:low_to_high_expr} showcases the comparison between the CNNs and QuNNs where the light blue line corresponds to CNNs and the other lines correspond to the QuNNs with different Ansatz circuits. As the intensity of attacks increased, both models experienced a decline in accuracy. However, the QuNNs demonstrated a more gradual decrease in performance. This trend was observed across various attack methods, including FGSM (Figure \ref{fig:low_to_high_expr} a, b on MNIST) and PGD (Figure \ref{fig:low_to_high_expr} c, d on FMNIST). The QuNNs consistently outperformed the CNNs in terms of accuracy for low and high adversarial conditions. This suggests that the QuNNs possess enhanced robustness against such attacks. 
Additionally, the research indicated that different QuNNs Ansatz variants displayed varying degrees of resilience to these adversarial challenges, which will be explored in detail in subsequent sections.

\subsection{Gradational Analysis of Expressibility and Entanglement}
\label{subsec:Low_to_high}
\begin{figure}[htb]
  \centering
  \includegraphics[width=0.47\textwidth]{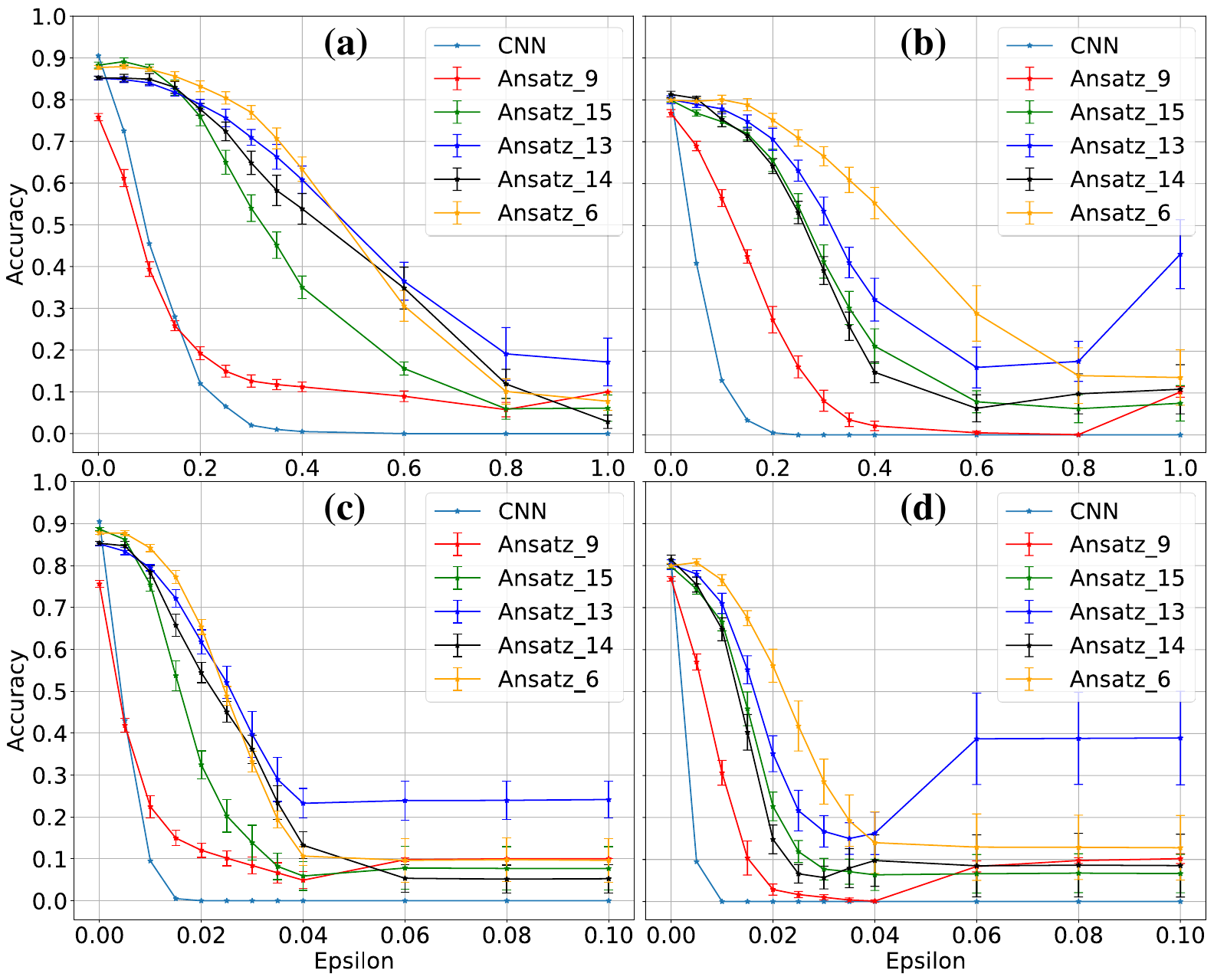}
  \caption{Low to high expressibility: (a) MNIST FGSM, (b) FMNIST FGSM, (c) MNIST PGD, (d) FMNIST PGD.}
  \label{fig:low_to_high_expr}
\end{figure}

\textbf{Low to high expressibility:} 
This experiment evaluates the resilience of QuNNs against adversarial attacks, specifically using the FGSM (Figure \ref{fig:low_to_high_expr} a-b) and PGD (Figure \ref{fig:low_to_high_expr} c-d). The attacks were conducted at various strengths across two datasets: MNIST (Figure \ref{fig:low_to_high_expr}a, c) and FMNIST (Figure \ref{fig:low_to_high_expr} b, d). The performance of QuNNs was analyzed by modifying the Ansatz in the quanvolutional filter. Five different Ansatz configurations were tested (Nos. 9, 15, 13, 14, and 6), selected based on their expressibility ranging from low to high, as detailed in Table \ref{table:ansatz_2}.
Results indicated that Ansatzes with higher expressibility generally enhanced the robustness of QuNNs against both types of attacks, particularly at lower perturbation strengths. Notably, despite having lower expressibility than Ansatz 14, Ansatz 13 exhibited superior performance. This result will be further explored in the section dedicated to Gate Selection Experiments. For the FMNIST dataset, the performance of Ansatzes 15 and 14 was notably similar.
Additionally, during PGD attacks, the performance metrics tended to plateau at an epsilon value of 0.04. This stabilization is attributed to pixel saturation, where pixel values are clipped to the range [0,1] upon exceeding a perturbation magnitude of 1, thereby maintaining data normalization, and this is observed throughout all following experiments. This clipping mechanism prevents the introduction of anomalous behavior by keeping adversarial perturbations within the valid range of natural images.

\begin{figure}[htb]
  \centering
  \includegraphics[width=0.47\textwidth]{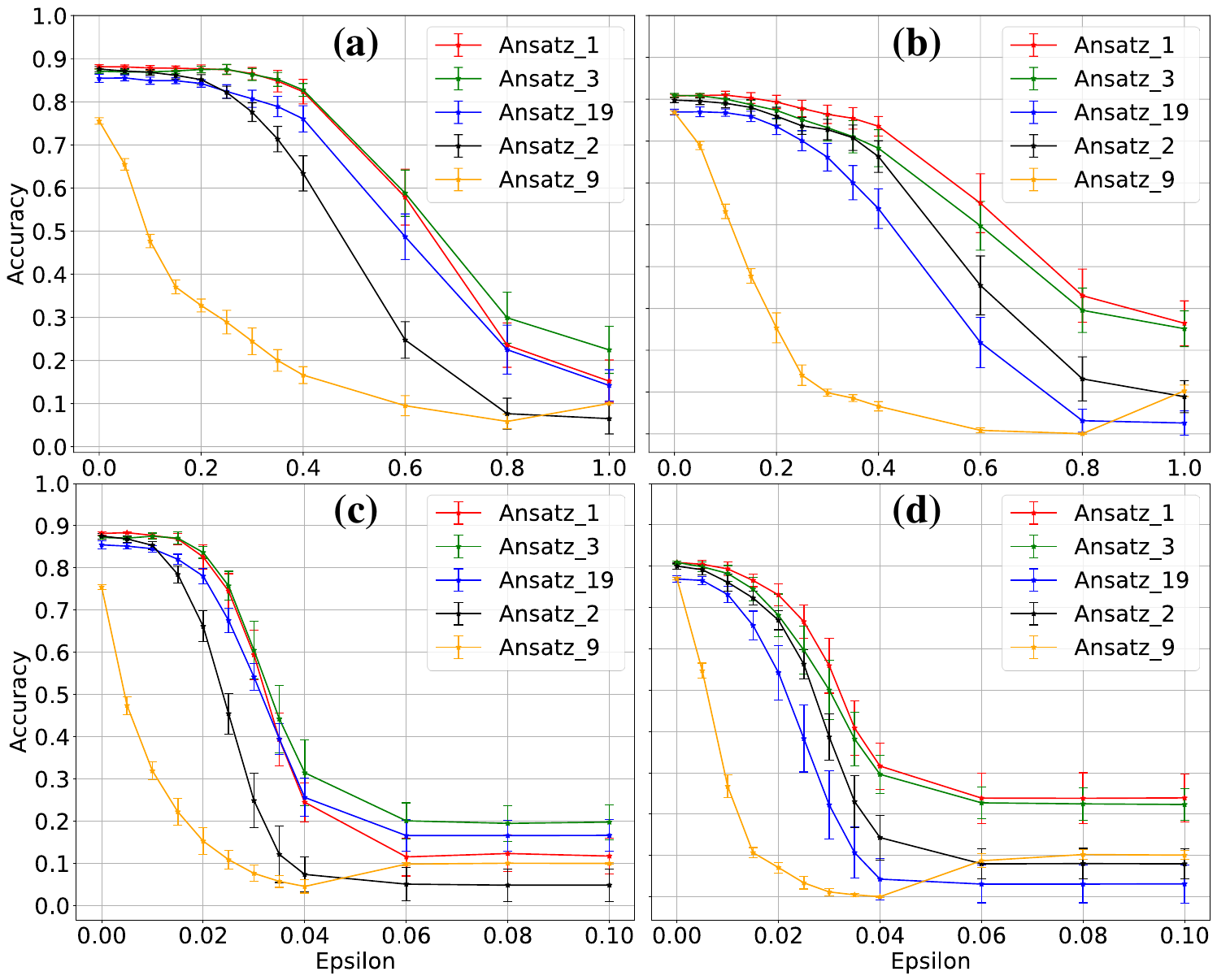}
  \caption{Low to high entanglement: (a) MNIST FGSM, (b) FMNIST FGSM, (c) MNIST PGD, (d) FMNIST PGD.}
  \label{fig:low_to_high_ent}
\end{figure}

\textbf{Low to high entanglement capability:} This experiment extends the methodology of the preceding evaluation (as demonstrated in Figure \ref{fig:low_to_high_ent}: FGSM attacks on MNIST and FMNIST (panels a and b), and PGD attacks on MNIST and FMNIST (panels c and d)), focusing on the variation of Ansatzes according to their entanglement capabilities. Five Ansatzes (Nos. 1, 3, 19, 2, and 9) were selected, showcasing a spectrum from low to high entanglement capabilities as outlined in Table \ref{table:ansatz_2}. Contrary to expressibility, the results demonstrated that Ansatzes with lower entanglement capabilities generally increased the robustness of QuNNs against both types of adversarial attacks across all tested perturbation strengths on both datasets.

Overall, these results establish a direct proportionality between expressibility and the robustness of QuNNs, as well as an inverse proportionality between entanglement and QuNNs robustness.

\subsection{Interdependency of Expressibility and Entanglement}
\label{subsec:Low_high_combination}

Given that the existing circuits do not align with our methodological requirements, we carefully selected Ansatz configurations that more closely matched our desired metrics. Specifically, for configurations requiring low entanglement and low expressibility, we selected Ansatz 1, which exhibits metrics indicative of low entanglement and medium expressibility. For cases where high entanglement and low expressibility were needed, Ansatz 9 was chosen as it satisfies these criteria. When low entanglement and high expressibility were required, Ansatz 3 was selected, as it presents low entanglement with medium expressibility, albeit higher than that of Ansatz 1. For scenarios necessitating both high entanglement and high expressibility, Ansatz 6 was selected due to its high entanglement (slightly above the medium threshold of 0.5) and high expressibility. The specific numerical values for these characteristics are detailed in Table \ref{table:ansatz_2}.

\begin{figure}[htb]
  \centering
  \includegraphics[width=0.47\textwidth]{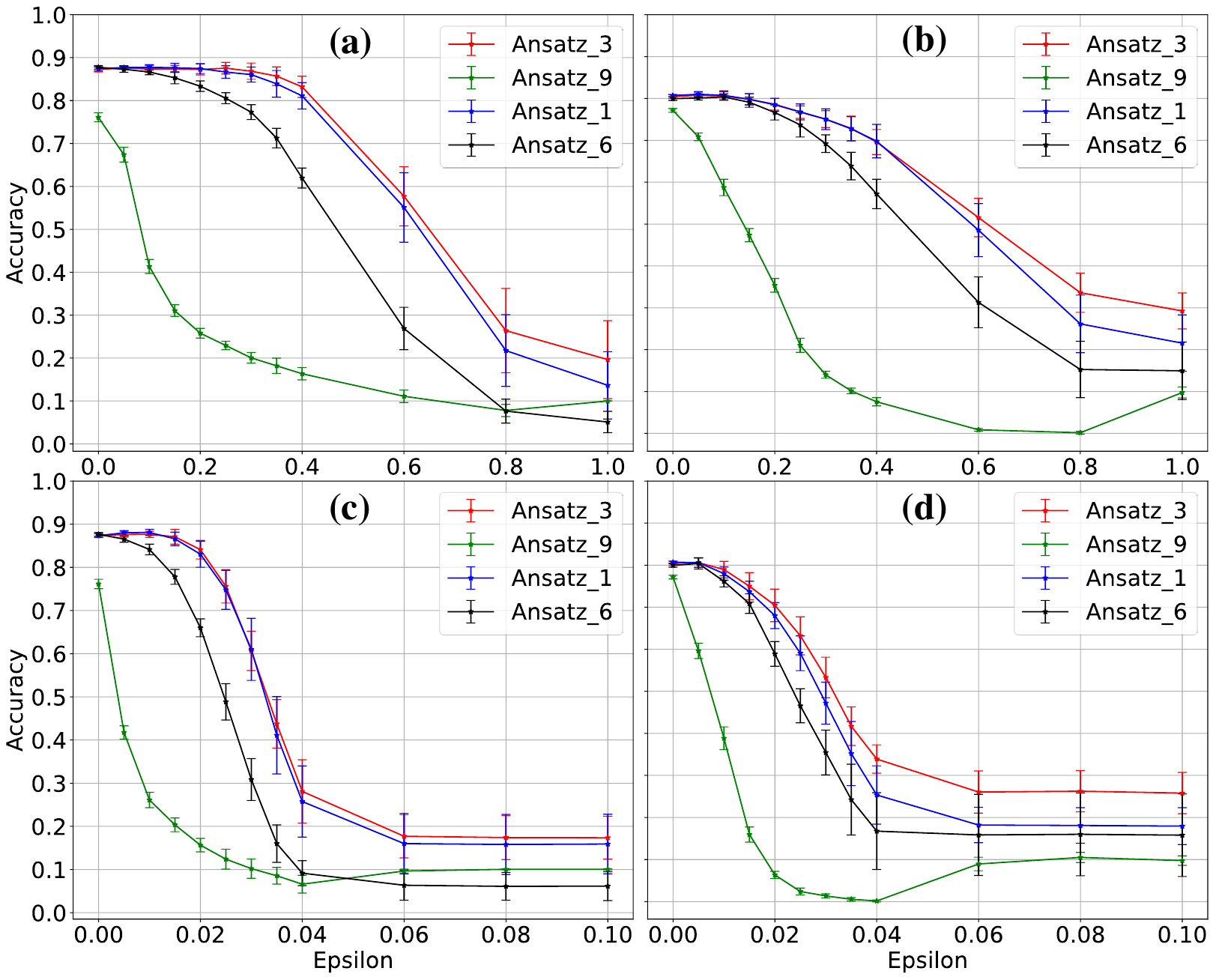}
  \caption{Low high combination: (a) MNIST FGSM, (b) FMNIST FGSM, (c) MNIST PGD, (d) FMNIST PGD.}
  \label{fig:low_high_combine}
\end{figure}

Figure \ref{fig:low_high_combine} illustrates the relationship between accuracy and epsilon for different attacks and datasets. Panels a and b show FGSM attacks on MNIST and FMNIST datasets, respectively, and panels c and d depict PGD attacks on MNIST and FMNIST datasets, respectively. When subjected to FGSM and PGD attacks on the MNIST and FMNIST datasets, Ansatz 9 demonstrated the lowest robustness, likely due to its combination of high entanglement and low expressibility, which appears to diminish its overall capability. Conversely, Ansatz 3 exhibited the best performance due to its balance of low entanglement and sufficient expressibility, which together enhance its robustness. Ansatz 1, while closely similar to Ansatz 3 in terms of low entanglement, exhibited slightly lower performance. This difference is attributed to the slightly reduced expressibility of Ansatz 1 (0.4015) compared to Ansatz 3 (0.2536), which affected its overall performance. On the other hand, Ansatz 6, while possessing high expressibility, was hindered by its high entanglement, leading to a degradation in performance. 

\subsection{Gate Selection Impact}
\label{gate_selection}

\begin{figure*}[htb]
  \centering
  \includegraphics[width=0.97\textwidth]{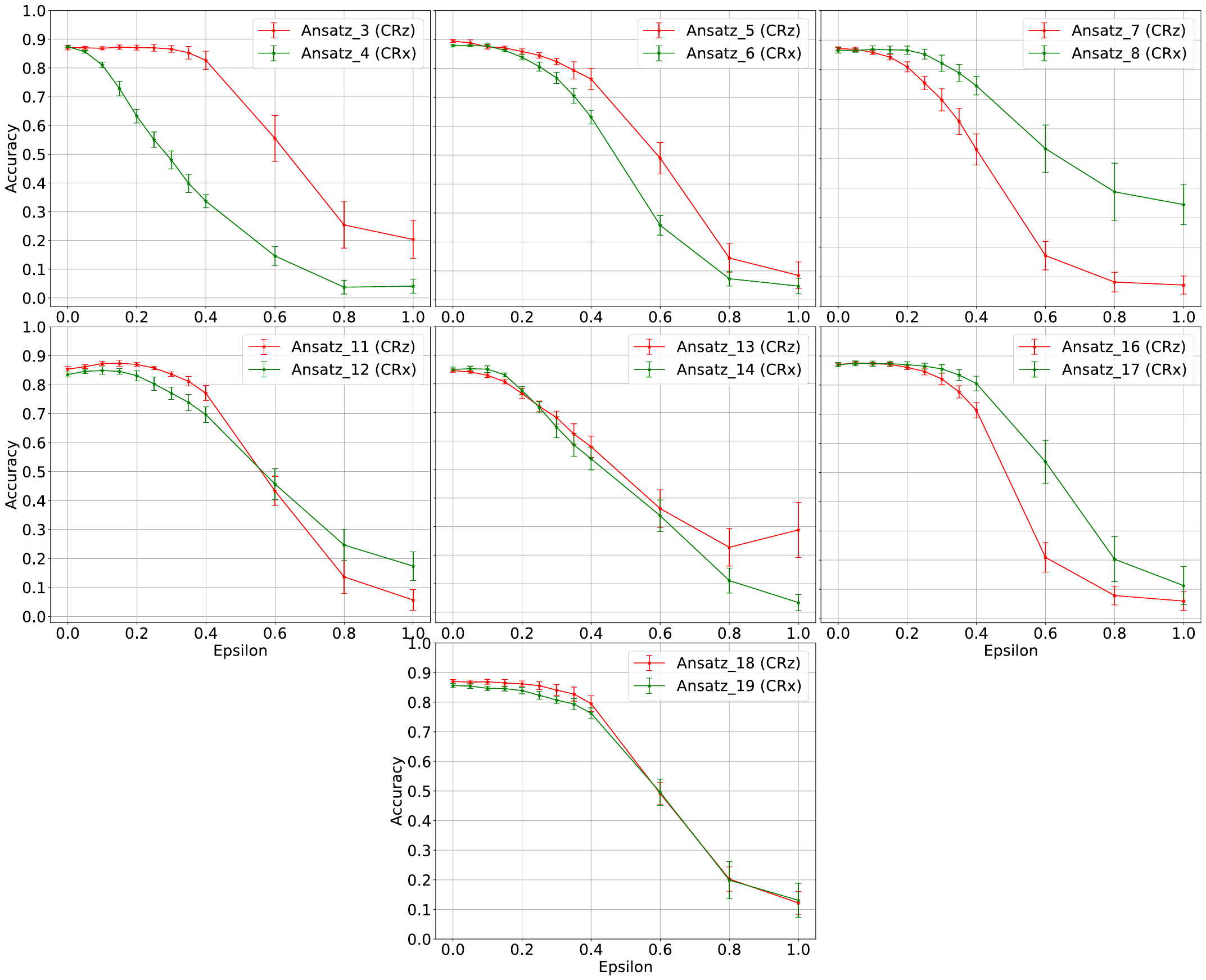}
  \caption{Impact of controlled rotation gate selection in the Ansatz on QuNN robustness. A Comparison Between $CRz$ and $CRx$ Gates for MNIST dataset under FGSM Attack.}
  \label{fig:gate_selection_MNIST}
\end{figure*}

\begin{figure*}[htb]
  \centering
  \includegraphics[width=0.97\textwidth]{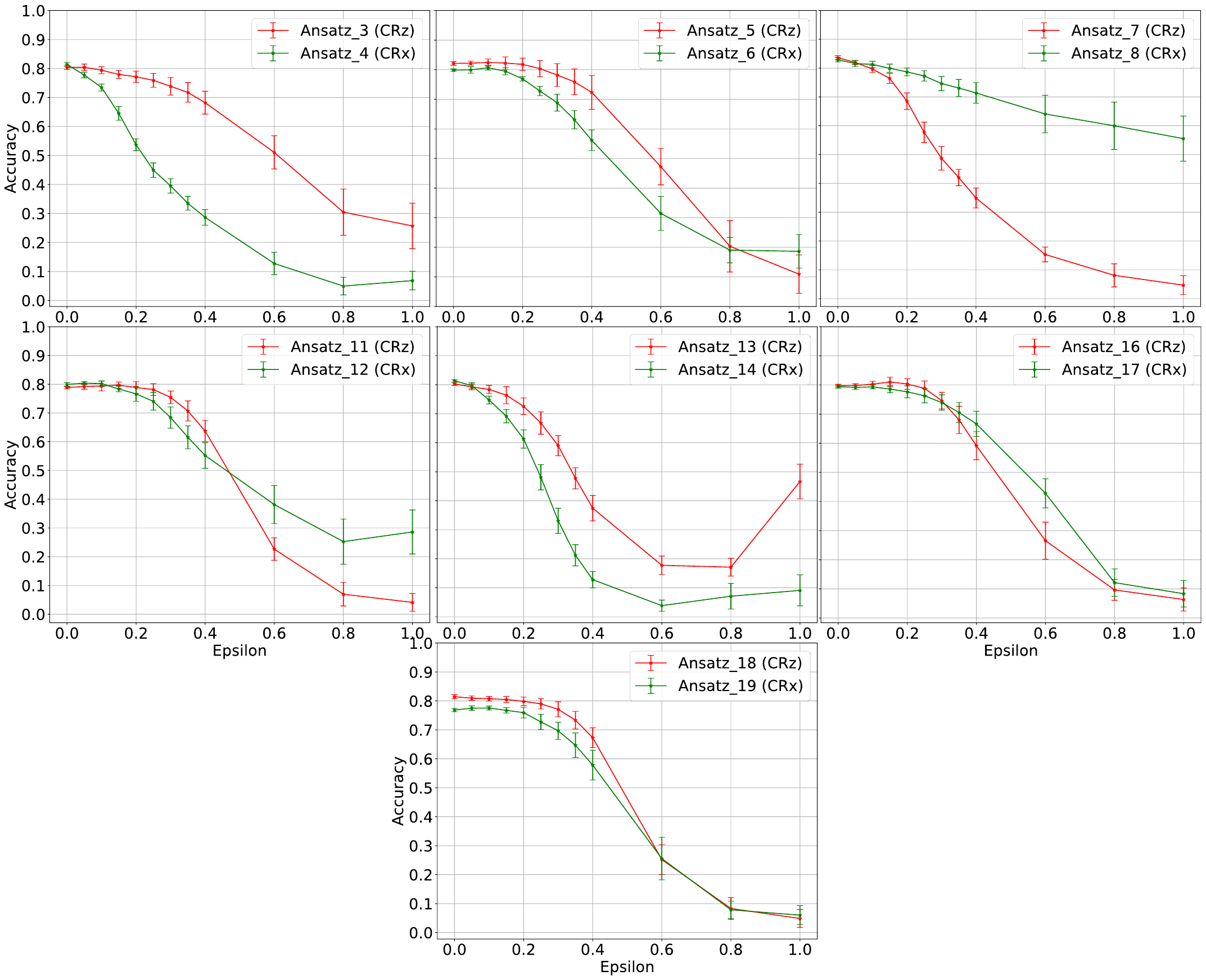}
  \caption{Impact of controlled rotation gate selection in the Ansatz on QuNN robustness. A Comparison Between $CRz$ and $CRx$ Gates for FMNIST dataset under FGSM Attack.}
  \label{fig:gate_selection_FMNIST}
\end{figure*}

\begin{figure*}[h!]
  \centering
  \includegraphics[width=0.88\textwidth]{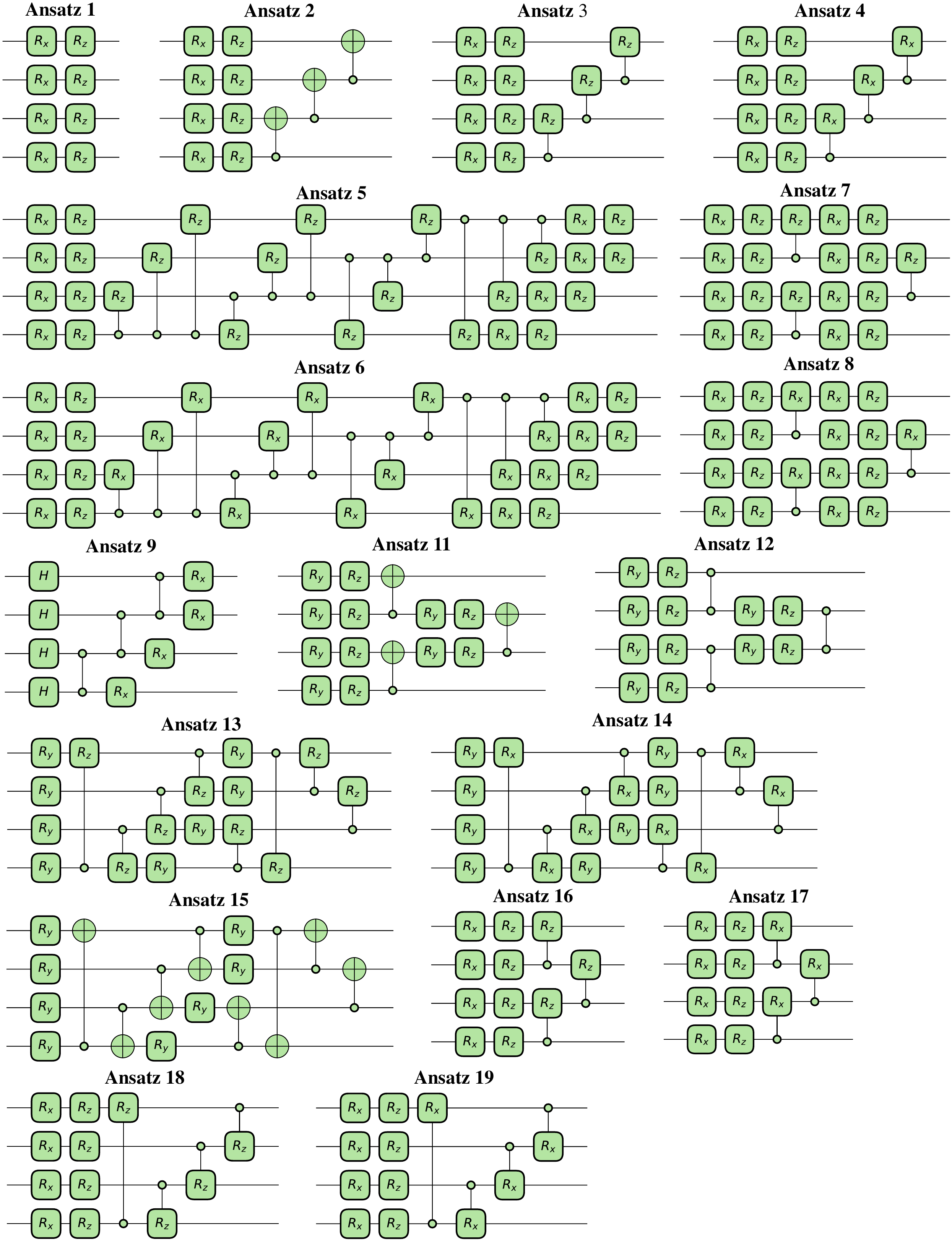}
  \caption{Template of Ansatzes, adapted from the referenced paper \cite{sim2019expressibility}}
  \label{fig:gate_selection}
\end{figure*}

In this experiment, we aimed to investigate methods to improve the robustness of QuNNs by examining the impact of different controlled rotation gates in the Ansatz structure. The Ansatzes evaluated in this study comprised pairs of quantum circuits, with one circuit in each pair utilizing a controlled rotation gate around the $Z$ axis ($CR_z$) and the other utilizing a controlled rotation gate around the $X$ axis ($CR_x$). Figure \ref{fig:gate_selection} presents a comparative analysis of the robustness of each Ansatz pair when subjected to the FGSM attack using the MNIST and FMNIST datasets.
In Figure \ref{fig:gate_selection_MNIST} and \ref{fig:gate_selection_FMNIST}, our findings indicate that Ansatzes incorporating  $CR_z$ in their entanglement operations generally exhibited enhanced robustness compared to those utilizing  $CR_x$, underscoring the significance of gate selection in the resilience of QuNNs in adversarial environments. Notably, in pairs 7-8 and 16-17, the Ansatz featuring  $CR_x$ demonstrated superior performance, which may be attributed to their increased expressibility.
The observed superior performance of Ansatz 13 compared to Ansatz 14 in section \ref{subsec:Low_to_high}, despite the higher expressibility of the latter, can be attributed to the specific quantum gates utilized in each Ansatz. Notably, Ansatz 13 incorporates the $CR_z$ gate, whereas Ansatz 14 employs the $CR_x$ gate. This distinction in gate selection is likely a key factor contributing to the enhanced performance of Ansatz 13.

\subsection{Impact of Quantum Circuit Design on Gradient Magnitudes}
To investigate why QuNNs exhibit superior adversarial robustness compared to CNNs, we measured the Average Gradient Magnitude (AGM) of the loss function with respect to input changes for both models using the MNIST dataset. The magnitude of gradients significantly impacts a model's vulnerability to adversarial attacks. High gradient magnitudes make the loss function highly sensitive to small input changes, allowing minor perturbations to cause significant loss variations. This increases the effectiveness of adversarial attacks by easily pushing inputs across decision boundaries. Conversely, low gradient magnitudes reduce the loss function’s sensitivity, making attacks harder. 

Table \ref{tab:agm} summarizes the AGM for the CNNs and several QuNNs Ansatz configurations. For illustration, we focus on the configurations tested in Figure \ref{fig:low_high_combine} (specifically, Ansatz 3, 9, 1, and 6). As shown in Table \ref{tab:agm}, QuNNs configurations with lower AGM values—such as Ansatz 3 and 1, which have AGMs of 0.7650 and 0.8249 respectively—exhibit significantly higher resistance to adversarial attacks compared to both the CNN (AGM = 1.7965) and high-AGM QuNNs variants (e.g., Ansatz 9 with an AGM of 3.8145). This finding aligns with the robustness hierarchy observed in our adversarial experiments (Fig. \ref{fig:low_high_combine}), where Ansatz 3 and 1 outperformed all other configurations. Although entanglement is an important aspect of quantum advantage, our data show that balancing entanglement with suitable expressibility, and thereby controlling AGM, is integral to achieving robust performance.

\begin{table}[h]
    \centering
    
    \begin{tabular}{lc}
        \hline
        \textbf{Method} & \textbf{Average Gradient Magnitude} \\
        \hline
        CNN & 1.7965 \\
        QuNN (Ansatz 3) & 0.7650 \\
        QuNN (Ansatz 9) & 3.8145 \\
        QuNN (Ansatz 1) & 0.8249 \\
        QuNN (Ansatz 6) & 1.5437 \\
        \hline
    \end{tabular}
    
    \caption{Average Gradient Magnitude for CNNs and various QuNNs Ansatz configurations.}
    \label{tab:agm}
\end{table}


\subsection{Discussion}

In this study, we have established that QuNNs display enhanced robustness compared to classical CNNs under white box attack scenarios using the FGSM and PGD methods on the MNIST and FMNIST datasets. This increased resilience can be attributed to the quanvolutional filters within QuNNs, which appear to function as protective mechanisms. These filters effectively mitigate perturbations by retaining meaningful and robust features of the images—features upon which the models were initially learned. In contrast, classical CNNs rely on informative features that, while useful, are more susceptible to adversarial attacks~\cite{west2023benchmarking},~\cite{west2023towards}. 

Our methodology, which evaluates three critical quantum circuit metrics—expressibility, entanglement capability, and controlled gate selection—has illuminated the substantial impact of Ansatz circuit characteristics on the robustness of QuNNs. We have demonstrated that highly expressive Ansatz designs enrich the representational capacity within the Hilbert space, endowing it with unique properties that help find robust representations of the data.
A more expressive Ansatz can explore a larger portion of the Hilbert space, allowing the quantum circuit to represent a wider variety of quantum states. This increased capacity enables the model to capture complex data structures and patterns, enhancing its ability to generalize and maintain performance even when faced with adversarial perturbations. 
This enhancement significantly bolsters the resilience of the QuNNs model against adversarial attacks.

Furthermore, our findings reveal that the entanglement capability of the quantum circuits exerts a negative impact on the effectiveness of expressive Ansatz designs, irrespective of their expressibility. Specifically, quantum circuits with high entanglement capability can generate highly entangled states, which are notably sensitive to external disturbances. This is likely due to the fact that any perturbation affecting one part of the system can influence the entire entangled state due to the strong correlations between qubits. In an adversarial setting, even small perturbations in the input data are distributed across the correlated system, amplifying their effect and leading to significant changes in the quantum state's amplitude and phase. As a result, these amplified effects degrade the robustness of the model under adversarial attacks, making it more susceptible to such perturbations.
Conversely, quantum states with lower degrees of entanglement might be less susceptible to such perturbations, leading to improved resilience of QuNNs against sophisticated adversarial inputs. This intriguing result suggests that there is a trade-off between entanglement capability and robustness, and that optimizing entanglement levels within quantum circuits could be a key factor in enhancing model robustness.

\begin{figure*}[htb]
  \centering
  \includegraphics[width=0.6\textwidth]{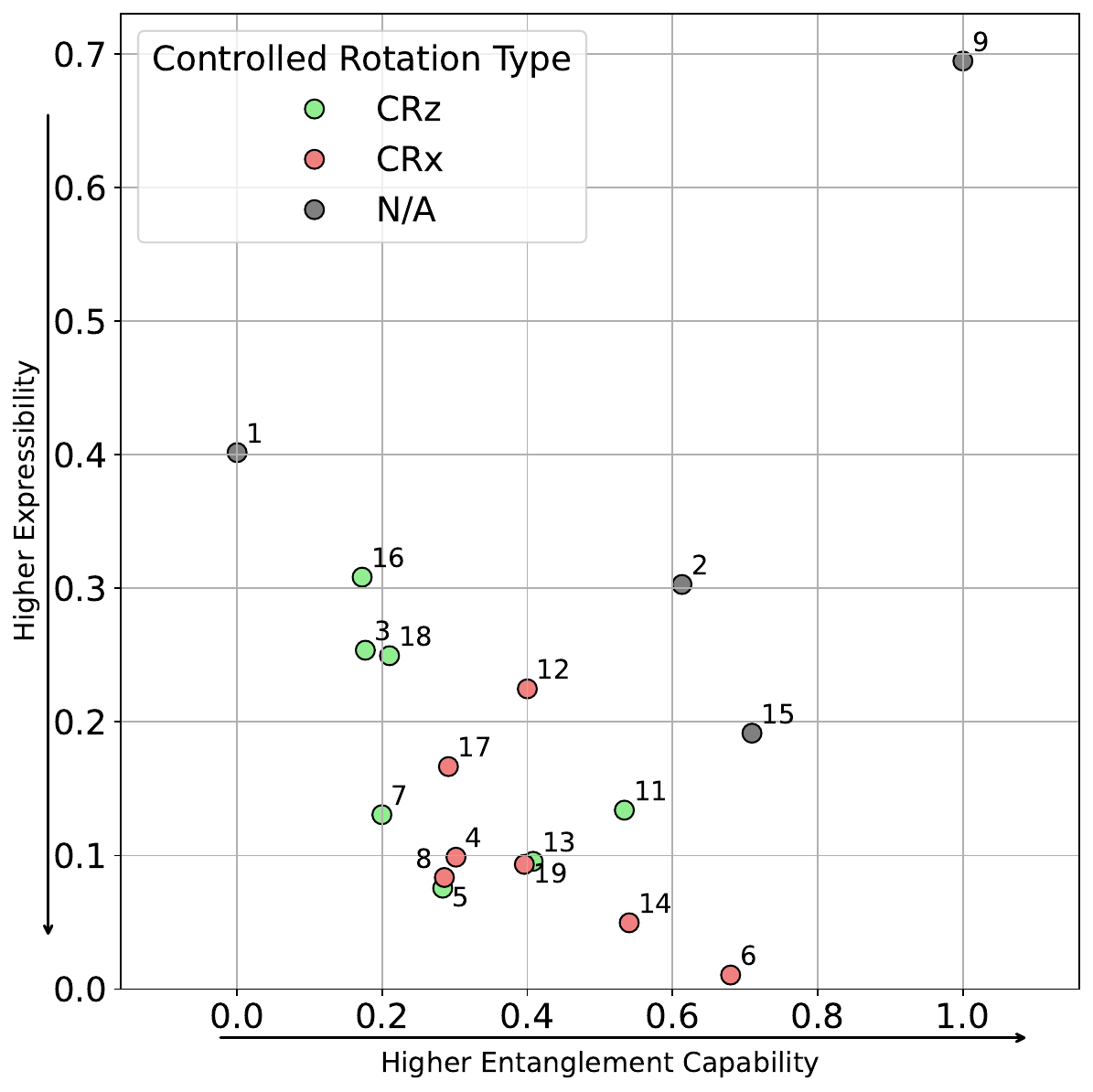}
  \caption{The plot shows Higher Expressibility vs. Higher Entanglement Capability for different Ansatzes, color-coded by rotation type (CRz, CRx, N/A). Numbers label each Ansatz. "N/A" indicates Ansatzes without paired counterparts for comparison in terms of rotation gate selection.}
  \label{fig:Spectrum}
\end{figure*}

An interesting finding in our work, contrary to the results presented in~\cite{hubregtsen2021evaluation}, which indicate that Ansatzes with controlled rotation gates around the $X$-axis $\left(CR_x\right)$ demonstrate strong accuracy performance on clean data. Our study reveals that, in adversarial environments, these $CR_x$ Ansatzes exhibit reduced robustness compared to those with controlled rotation gates around the $Z$-axis ($CR_z$).
Furthermore, as discussed in~\cite{sim2019expressibility}, Ansatzes incorporating $CR_x$ gates tends to enhance the expressibility of quantum circuits. While our results confirm that more expressive Ansatzes are generally more robust, our findings also demonstrate that Ansatzes with $C R_z$ gates, despite their lower expressibility, offer superior robustness. 
This disparity can be attributed to the fact that CRx gates generally increase the entanglement capability of the Ansatz circuits, as shown in Figure \ref{fig:Spectrum}. High entanglement capability means that perturbations affecting one qubit can influence the entire system due to strong correlations between qubits, amplifying the impact of adversarial perturbations and making the model more susceptible to attacks.  In contrast, $CR_z$ gates primarily affect the phase relationships between qubits, rather than both amplitude and phase as $CR_x$ gates do. This controlled entanglement localizes the impact of adversarial inputs, reducing the spread of perturbations and enhancing the model's robustness.
This suggests that gate selection plays a critical role in enhancing the robustness of QuNNs, and that expressibility should not be considered the sole metric in this context.

The model's performance under the PGD attack exhibits unexpected deviation in certain cases when the perturbation strength exceeds a value of 1. This anomaly occurs because, at such high levels of perturbation, the altered image ceases to be meaningful to both humans and machines. The PGD attack operates iteratively, incrementally increasing the perturbation with each iteration. Typically, the process begins with a perturbation of 0.06 and progressively escalates towards a perturbation of 1. Once this threshold is crossed, the images become highly distorted, leading to erratic and unpredictable model performance. This deterioration in performance can be attributed to the fact that the input images, beyond a certain level of perturbation, no longer resemble their original forms, thus confusing the model and impairing its ability to function effectively.

Our study offers valuable insights into the adversarial robustness of QuNNs, but several limitations should be acknowledged. First, we kept the convolutional layers non-trainable 
to act as a fixed filter in the quanvolutional layer. While this approach served our goal, it constrained the model from fully utilizing its parameters during training. Future research could investigate the effect of making these layers trainable to provide a more comprehensive assessment of model robustness, considering the computational resources and time required. 
Second, the results presented in this paper are generated using a simulator, employing actual quantum processors will allow us to validate our findings in a real quantum environment. However, due to the substantial resources and time required to execute a large number of quantum circuits, specifically, for $1000$ images, each consisting of $14 \times 14$ patches, resulting in $196$ circuits per image, the total number of circuits would amount to $196000$. Additionally, hardware limitations and the queueing time for accessing quantum computing resources further complicate this process. Consequently, the full implementation of this approach will be considered as part of future work.
Lastly, we noticed some exceptions in our results, such as circuits using $CRx$ gates outperforming those with $CRz$ gates. These exceptions suggest that additional factors may influence robustness. Further exploration of these cases is needed to uncover the underlying mechanisms and identify other contributors to the adversarial robustness of QuNNs.
    
Our experiment demonstrates that the architecture of quantum circuits plays a pivotal role in determining model robustness. Notably, Ansatzes with higher expressibility appear to provide increased resistance to white-box attacks, particularly when they incorporate low entanglement. This relationship is underscored by our findings on AGM, where robust architectures consistently display lower AGM scores. We argue that highly expressive circuits yield complex mappings from input to feature space, resulting in irregular loss landscapes with numerous local minima and convoluted gradient structures. Consequently, standard gradient-based attacks struggle to find reliable descent directions, underscoring the significance of circuit design in achieving robustness. These observations align with recent studies showing that increasing the complexity of the gradient landscape, even via classically inspired methods such as noise injection \cite{he2019parametric} or gradient masking \cite{lee2020gradient}, can enhance resistance against white-box adversaries. Our contribution expands this concept into the quantum domain, harnessing phenomena such as entanglement and expressibility to enhance gradient complexity and strengthen resilience against adversarial attacks.

We acknowledge that lower gradients may result from gradient masking, which could potentially obscure underlying vulnerabilities. While lower gradients indicate stability, they might leave the model susceptible to advanced attacks, such as black-box~\cite{el2024robqunns} or iterative methods. To address these limitations, future research will focus on developing more advanced defense protocols and regularization techniques to ensure robustness against a broader range of attack scenarios.


\section{Conclusion}

In this study, we developed a comprehensive methodology to examine the interplay between three key metrics of the Ansatz circuit in quanvolutional filters: expressibility, entanglement capability, and gate selection. Our experiments, conducted using the MNIST and FMNIST datasets, evaluated the performance of these filters under adversarial conditions facilitated by two white-box attack methods: FGSM and PGD.
Initially, our results underscored a quantum advantage by demonstrating enhanced robustness in QuNNs compared to their classical counterparts. We then advanced our investigation by systematically testing various Ansatz designs, ranging from low to high expressibility, and similarly assessing variations in entanglement capability. Subsequent analyses focused on the reciprocal influence of expressibility and entanglement, providing new insights into their interconnected dynamics.
Finally, our study delved into the effects of different controlled rotation gate selections on the robustness of the networks, revealing significant implications for the design of more resilient quantum computing architectures. This novel approach not only highlights the potential of QuNNs in adversarial environments but also sets the groundwork for future explorations into optimizing quantum circuits for machine learning applications.

Building on our methodology, this study established a critical relationship between expressibility, entanglement capability, and gate selection, and their collective impact on the robustness of quantum classifiers against adversarial attacks. We identified a direct correlation between the capacity of an Ansatz to uniformly explore the Hilbert space—an attribute of highly expressive Ansatzes—and the enhancement of the overall robustness of the system. Furthermore, our findings underscore the significance of gate selection, particularly highlighting the role of Ansatzes featuring controlled rotation gates around the Z-axis. This insight is pivotal for the design of robust quantum circuits and should be a key consideration in the development of resilient architectures for quantum classifiers. Additionally, our analysis showed that highly expressive circuits lead to low AGM, which suggests a complex loss landscape, making it harder for gradient-based attacks to find consistent descent directions. These conclusions not only advance our understanding of quantum computing resilience but also suggest practical guidelines for optimizing quantum classifier performance in the presence of adversarial threats. 

Our research has demonstrated that QML models not only promise computational speed but also exhibit additional robustness advantages. To enhance the scope of our work, future endeavors will focus on exploring a broader array of quantum circuits. We aim to develop methodologies for designing quantum circuits that are tailored to achieve specific quantum metrics values, optimizing architecture for robust performance against adversarial threats.
Moreover, we plan to extend our findings by applying our methodologies to a wider variety of QML models, across diverse datasets and under various attack scenarios, while incorporating different encoding techniques to evaluate data representation impact on robustness. This strategic expansion will pave the way for robustifying QML systems through the custom design of quantum circuit architectures. Such efforts represent a significant shift towards embedding inherent robustness directly into the structural fabric of QML models, markedly enhancing their resilience against adversarial threats. This innovative approach promises to fortify the reliability and security of quantum computing applications in machine learning.

\begin{acks}
This work was supported in parts by the NYUAD Center for Cyber Security (CCS), funded by Tamkeen under the NYUAD Research Institute Award G1104, and the NYUAD Center for Quantum and Topological Systems (CQTS), funded by Tamkeen under the NYUAD Research Institute grant CG008.
\end{acks}

\bibliographystyle{ieeetr}
\bibliography{main.bib}

\onecolumn

\section*{Appendix A: Ansatz circuits} \label{appendixA}

\end{document}